\documentclass[aps,prd,a4paper,preprintnumbers,amsmath,amssymb,showpacs,reprint]{revtex4-1}
\usepackage{amssymb}
\usepackage{slashed}
\usepackage{hyperref}

\newcommand{\Mp}{M_{\mathrm{P}}}

\DeclareMathOperator{\tr}{tr}
\DeclareMathOperator{\Tr}{Tr}

\renewcommand{\d}{\mathrm{d}}
\newcommand{\D}{D}
\newcommand{\e}[1]{\mathrm{e}^{{#1}}}
\newcommand{\im}{\mathrm{i}}
\newcommand{\Lag}{\mathcal{L}}
\newcommand{\LagE}{\Lag_E}
\newcommand{\Nf}{N_{\mathrm{f}}}

\begin{document}

\title{Anomalous coupling of scalars to gauge fields}

\author{Philippe Brax}
\affiliation{Institut de Physique Th\'eorique, CEA, IPhT, CNRS, URA 2306,
  F-91191Gif/Yvette Cedex, France}
\email{philippe.brax@cea.fr}
\preprint{DESY 10-167}

\author{Clare Burrage}
\affiliation{D\'{e}partment de Physique Th\'{e}orique,
Universit\'{e} de Gen\`{e}ve, 24 Quai E. Ansermet,
CH-1211, Gen\`{e}ve, Switzerland}
\affiliation{Theory Group, Deutsches Elektronen-Synchrotron DESY,
D-22603, Hamburg, Germany}
\email{Clare.Burrage@unige.ch}

\author{Anne-Christine Davis}
\affiliation{Department of Applied Mathematics and
Theoretical Physics, Centre for Mathematical Sciences, Cambridge CB3
0WA, United Kingdom}
\email{A.C.Davis@damtp.cam.ac.uk}

\author{David Seery}
\affiliation{Department of Physics and Astronomy, University of
Sussex, Brighton, BN1 9QH, United Kingdom}
\email{D.Seery@sussex.ac.uk}

\author{Amanda Weltman}
\affiliation{Astronomy, Cosmology and Gravity Centre, Department of Mathematics and Applied Mathematics,
University of Cape Town, Private Bag, Rondebosch, South Africa, 7700}
\email{amanda.weltman@uct.ac.za}

\begin{abstract}
	We study the transformation properties of a scalar--tensor theory,
	coupled to fermions, under
	the Weyl rescaling associated with a transition from the
	Jordan to the Einstein frame.
	We give a simple derivation of the corresponding modification to
	the gauge couplings.
	After changing frames, this gives rise to a
	direct coupling between the scalar and the gauge fields.
\end{abstract}

\maketitle

\section{Introduction}
Axions have a coupling to gauge fields resulting from the existence
of a chiral anomaly. Similarly, string axions must couple to gauge fields
in order to cancel gauge anomalies using a four-dimensional version of
the Green--Schwarz mechanism \cite{Witten:1985fp, *Segre:1986zk}.
These gauge interactions of axions and
axion-like particles
leads to a rich phenomenology.
They give rise to important effects in astrophysics, causing
stars to cool at an accelerated rate \cite{1992ApJ, *Raffelt:1994ry,
*Raffelt:2006cw}.
At another extreme there are interesting
optical phenomena, leading to a prospect of detecting
axion-like particles with cavity-based
experiments such as ALPS and GammeV
\cite{Ehret:2009sq, *Ehret:2010mh, *Chou:2008gr, *Upadhye:2009iv,
*Steffen:2010ze},
or experiments such as CAST and
the Tokyo axion helioscope
\cite{Zioutas:2004hi, *Andriamonje:2007ew, *Inoue:2008zp, *Inoue:2010qm}.

Scalar fields have attracted considerable attention following the discovery,
less than fifteen years ago, that the
expansion rate of the universe is accelerating
\cite{Riess:1998cb, *Perlmutter:1998np}.
It has long been known that scalar fields can play a significant role
in a phase of early-universe accelerated expansion, or `inflation.'
The same is true in the late universe. However, if a scalar `dark energy'
field is responsible for the acceleration measured today, there is
difficulty. To modify the expansion rate on cosmological
scales, the mass of the field should be as low as the Hubble rate today,
of order $10^{-33} \; \text{eV}$. But this would lead to the existence
of a `fifth force,' violating the weak equivalence principle.
Fortunately, mechanisms making the small mass of the scalar
field phenomenologically acceptable
on large scales---and preventing the appearance of a
fifth force in the solar system and in the laboratory---have been discovered.

In certain dark energy models, acceleration is due to the rolling
of a scalar field over a runaway potential
\cite{Damour:1992kf, *Damour:1993id,
Damour:1994zq, *Damour:1994ya, Veneziano:2001ah,
*Damour:2002mi, *Damour:2002nv, Kachru:2003aw, *Kachru:2003sx}.
Such potentials have a long history in attempts to explain the
dark energy scale \cite{Zlatev:1998tr}.
Two methods have been invoked to eliminate the unwanted scalar
fifth forces which accompany them: (a) either the \emph{chameleon mechanism}
\cite{Khoury:2003aq, *Khoury:2003rn} or the
\emph{Vainshtein effect}~\cite{Vainshtein:1972sx}, in which the scalar
field is screened by massive bodies due to non-linear effects;
or (b) the \emph{Damour--Polyakov} mechanism
\cite{Damour:1994zq, *Damour:1994ya}, where the coupling of the scalar
field to matter is constructed to vanish dynamically.
Similar effects can be achieved in models of modified gravity.
Indeed, under certain circumstances,
modified gravities such as the Dvali--Gabadadze--Poratti model
\cite{Dvali:2000hr, *Luty:2003vm, *Porrati:2004yi},
$f(R)$ theories
\cite{Carroll:2003wy, *Capozziello:2003tk},
or the Galileon model
\cite{Nicolis:2008in, *deRham:2010eu}
reduce to the inclusion of a new, scalar degree of freedom,
whose fifth force is either screened by a chameleon mechanism
\cite{Brax:2008hh, *Hu:2007nk}
or the Vainshtein effect. In all these
theories the dynamics of the scalar field are determined by a
scalar--tensor theory.

Does the scalar
field $\phi$ couple to gauge fields such as the photon?
If so, the rich optical phenomenology of axion-like particles
could be extended to models of dark energy or modified gravity
\cite{Brax:2007ak,
*Brax:2007hi, *Burrage:2007ew, *Burrage:2008ii, *Burrage:2009mj,
*Schelpe:2010he, Brax:2010jk, Davis:2009vk, *Davis:2010nj}.
Since $\phi$ is a gauge-singlet, this coupling must
involve an interaction with the gauge field kinetic term
of the form $f(\phi) F_{ab} F^{ab}$,
where $F_{ab} = \partial_{[a} A_{b]}$
is the field strength associated with the gauge field
$A_a$.

The possibility of such couplings was
investigated by Kaplunovsky \& Louis
\cite{Kaplunovsky:1994fg}
in the context of locally supersymmetric effective quantum field theories.
After Weyl rescaling,
Kaplunovsky \& Louis found that
a Wess--Zumino term was generated, associated with a super-Weyl anomaly,
which necessarily induced
an $f(\phi) F_{ab} F^{ab}$ interaction.
The function $f(\phi)$ was determined entirely at one-loop level.
Kaplunovsky \& Louis used these effects to derive exact threshold corrections
\cite{Ross:1978wt, *Hall:1980kf}
to the gauge couplings from integrating out species above the
supersymmetry breaking scale.
The study of such threshold corrections in string theories has generated
a large literature.

Kaplunovsky \& Louis' result was framed in the context
of supergravity.
Doran \& J\"{a}ckel later
observed that changing frame
would typically lead to a redefinition of couplings
in any effective field theory
\cite{Doran:2002bc}.
In this brief note we give an alternative, simpler derivation
of Kaplunovsky \& Louis' result
which does not
make use of the methods of supersymmetry or require the application
of superfield Feynman rules,
and which applies in any low energy effective theory.

\vspace{2mm}
\noindent
\emph{Weyl rescaling.}---%
A range of scalar--tensor theories exist.
Writing the theory in terms of a metric $\tilde{g}_{\mu\nu}$
defined so that the new scalar field does not directly interact with
matter---the `Jordan frame'---we focus on theories which can be
written in the form
\begin{equation}
\begin{split}
	S_J = & \int \d^4 x \; {\sqrt {-\tilde g}}
	\Bigg[
		\frac{\Mp^2}{2} B^2(\phi) \tilde{R}
		- \frac{1}{2} (\partial \phi)^2 -V(\phi)
	\Bigg] \\
	& \mbox{}
	+ S_m(\psi^i, \tilde g_{\mu\nu}) ,
\end{split}
\label{eq:action}
\end{equation}
where $\phi$ is the new scalar degree of freedom,
$B(\phi)$ is an arbitrary function,
$\Mp^2 = (8 \pi G)^{-1}$ is the reduced Planck mass, and
$\tilde R$ is the Ricci scalar constructed using the metric
$\tilde g_{\mu\nu}$. The matter action, $S_m$, involves an arbitrary
collection of fields $\psi^i$ coupled to $g_{\mu\nu}$, but not
$\phi$. Eq.~\eqref{eq:action} includes a wide class of dark energy
models, but does not include infrared modifications
of Einstein gravity such as the `Galileon'
\cite{Nicolis:2008in}.

In this frame, $\phi$ is coupled non-minimally to gravity via
$B^2(\phi) \tilde{R}$.
In the `Jordan frame,' this encapsulates
how Einstein gravity is modified.
However, it is possible to describe the same modification of gravity
in different ways,
by making field redefinitions which do not change the physics.
The Jordan frame action is classically equivalent to an `Einstein frame'
theory, written in terms of a redefined metric $g_{\mu\nu}$,
which satisfies
\begin{equation}
	g_{\mu\nu} = B^{-2}(\phi) \tilde{g}_{\mu\nu} .
	\label{eq:weyl-transform}
\end{equation}
This rescaling is an example of a Weyl transformation.
In the Einstein frame, the scalar field is minimally coupled to
$g_{\mu\nu}$ but couples directly to the matter fields $\psi_i$.
There is no principle which can tell us whether the Einstein
frame or Jordan frame is more fundamental.
Which we pick is simply a matter of obtaining the most
convenient description for the problem at hand.

\vspace{2mm}
\noindent
\emph{Field-dependent gauge couplings.}---%
It is well-known that the Maxwell kinetic term
$F_{ab} F^{ab}$ is invariant under Weyl rescalings.
Classically, therefore, it follows that
if this interaction is absent in one frame
it is absent in \emph{all} frames, and its inclusion
or otherwise
is merely a
free choice to be made in model-building.

Kaplunovsky \& Louis' result implies that,
after quantization, the change of variables
associated with shifting from one frame to another
naturally induces a coupling between $\phi$ and the kinetic term of any
gauge field, either Abelian or non-Abelian, which is coupled to fermion
species.
Therefore, quantization and change of frame do not commute.
From the standpoint of an effective field theory
it is unnatural to take the interaction to be absent:
this corresponds to picking one choice of frame as
`fundamental,' and all others as `derived.'
There is no justification for such a choice.
Instead, the coupling should be included
and its magnitude constrained by experiment.

In \S\ref{sec:anomaly} we make a quantitative estimate of the
effect, by
explicit calculation of the Jacobian associated with change of
variables in the path integral.
This can be accomplished using a
method introduced by Fujikawa to calculate the chiral anomaly
\cite{Fujikawa:1979ay, *Fujikawa:1980eg}.
A related calculation of the conformal anomaly in a chameleon model
with conformally coupled scalars was given by Nojiri \& Odintsov
\cite{Nojiri:2003ti}.
Similar anomalous Jacobians were encountered by
Arkani-Hamed \& Murayama \cite{ArkaniHamed:1997mj}
while studying exact $\beta$-functions in supersymmetric
gauge theories.
Arkani-Hamed \& Murayama referred to the appearance of a non-trivial
Jacobian as a ``rescaling anomaly.''
In \S\ref{sec:discuss} we summarize the calculation and discuss
our conclusions.

We work in a spacetime with signature $(-,+,+,+)$, and choose units
so that $\hbar = c = 1$.
Throughout, we suppress spinor indices.
The four-dimensional Dirac matrices are
$\gamma^a$, and satisfy the algebra
$[ \gamma^a , \gamma^b ] = 2 \eta^{ab}$.
With our sign conventions, the conjugate of a Dirac
spinor is $\bar{\lambda} = \lambda^\dag \gamma^0$. 

\section{Gauge couplings from Weyl transformations}
\label{sec:anomaly}

Our starting point is the scalar--tensor action,
Eq.~\eqref{eq:action}.
A redefinition of couplings arises on changing frame
because it is not possible to define the
path integral measure in a way which is invariant under
transformations of the form~\eqref{eq:weyl-transform}.
In our formalism, this is the analogue of the Weyl rescaling
studied by Kaplunovsky \& Louis,
given in Eq. (2.23) of their paper
\cite{Kaplunovsky:1994fg}.

To define the measure, we
formally compactify spacetime on a large manifold after performing a Wick rotation.
This is an intermediate step: our conclusions are independent of
the details of the compactification, and continue to apply if we
later revert to a noncompact space. To simplify the presentation, we have kept the Minkowski-convention signature throughout. 
We write the vielbein on this compact manifold as $e_\mu^a$;
Greek letters label spacetime indices, whereas
Latin letters label Lorentz indices associated with the tangent bundle.
The massless Dirac equation has a discrete eigenvalue
spectrum $\lambda_n$, where $n = 1, 2, \ldots$,
\begin{equation}
	(\slashed{e}^\mu \D_\mu) \psi_n = \lambda_n \psi_n ,
\end{equation}
where $\slashed{q} = \gamma^a q_a$
for any tangent-space vector $q_a$,
and $\psi_n$ is a Dirac spinor, obeying suitable boundary conditions
if necessary.
This step is formally valid only after Wick rotation to Euclidean
signature.

The derivative operator $\D_\mu$ includes
appropriate gauge terms for charged fermion species,
\begin{equation}
	\D_\mu = \partial_\mu - \im e A_\mu + \omega_\mu ,
\end{equation}
where $\omega_\mu$ is the spin connexion, $e$ is a coupling constant, and
$\omega_\mu = \frac{1}{8} [ \gamma_a , \gamma_b ] \omega_{\mu}^{ab}$.
We generally neglect this term in what follows, since it is
inessential
for obtention of
the anomalous scalar interactions with
gauge fields.
In any case,
where the compactification manifold is a flat torus, this gives
exact results.

The $\psi_n$ form a complete, orthogonal set of basis functions.
We choose a normalization so that
$\int \d^4 x \, \sqrt{-g} \, \bar{\psi}_m \psi_n = \delta_{nm}$.
In terms of this basis,
a generic massless spinor $\lambda_J$, or conjugate spinor $\bar{\lambda}_J$,
can be represented by a set of coefficients
$\{ a_n, \bar{b}_n \}$, where we have written
\begin{equation}
	\lambda_J = \sum_n a_n \psi_n
	\quad
	\text{and}
	\quad
	\bar{\lambda}_J = \sum_n \bar{b}_n \bar{\psi}_n .
\end{equation}
The subscript `$J$' denotes that these are Jordan-frame fields.
It is a familiar idea that
the measure $[ \d \lambda \; \d \bar{\lambda}]$
for integration over
$\lambda$ and $\bar{\lambda}$
can be defined by integration over the
coefficients $a_n$ and $\bar{b}_n$.
We define
\begin{equation}
	[ \d \lambda \; \d \bar{\lambda} ]_J
	= \prod_n M^3 \, \d a_n \, \d \bar{b}_n ,
	\label{eq:measure-def}
\end{equation}
where $M$ is mass scale needed to make the measure dimensionally
correct, but which plays little role in the analysis
and will not appear in subsequent expressions.
Eq.~\eqref{eq:measure-def}
expresses the functional integral in terms of
a measure on the Jordan-frame fields.
We are interested in determining the transformation law
connecting~\eqref{eq:measure-def} with
the same measure expressed in terms of Einstein-frame fields.
As we will see, this
typically contains local divergences which can be absorbed in a
redefinition of the Einstein frame Lagrangian.

On translation to the Einstein frame, the matter action acquires
interactions with $\phi$. In particular, the fermion kinetic term is
not conformally invariant and mixes with $\phi$. To obtain canonically
normalized Einstein-frame fields we make the change of variable
\begin{equation}
	\lambda_E = B^{3/2} \lambda_J
	\quad
	\text{and}
	\quad
	\bar{\lambda}_E = B^{3/2} \bar{\lambda}_J ,
\end{equation}
where we have used that $B$ is real.
The S-matrix is invariant under field redefinitions, so
this transformation does not change the physical content of the theory.
Written in terms of these fields,
the Einstein frame Lagrangian for each species of fermion satisfies
\begin{equation}
	\LagE \supseteq - \bar{\lambda}_E (\slashed{e}^\mu \D_\mu) \lambda_E
	+ \text{h.c} ,
\end{equation}
where `h.c.' denotes the Hermitian conjugate of the preceding term.

This field redefinition is associated with a Jacobian,
or change of measure,
represented by a fermionic determinant.
To calculate it, we represent the Einstein-frame spinor fields
in terms of the basis functions $\psi_n$ and $\bar{\psi}_n$ with
coefficients $c_n$ and $\bar{d}_n$.
Using~\eqref{eq:measure-def}, the change of variables is
\begin{equation}
	[ \d \lambda \; \d \bar{\lambda} ]_J =
	\left| \frac{\partial( a, \bar{b} )}
		{\partial( c, \bar{d} )}
	\right|
	[ \d \lambda \; \d \bar{\lambda} ]_E .
\end{equation}
where the Jacobian determinant satisfies
\begin{align}
	\nonumber
	& \left| \frac{\partial( a, \bar{b} )}
		{\partial( c, \bar{d} )}
	\right|
	=
	\left( \det \frac{\partial a_k}{\partial c_\ell}
		\det \frac{\partial \bar{b}_m}{\partial \bar{d}_n}
	\right)^{-1} \\
	& \quad = \exp \tr \ln \int \d^4 x \, \sqrt{-g} \;
		B^{3/2} ( \bar{\psi}_m \psi_n + \bar{\psi}_n \psi_m ) ,
	\label{eq:jacobian}
\end{align}
where the trace is over the basis indices $m$ and $n$.

The completeness relation for $\psi_n$ and $\bar{\psi}_n$
reads $\tr \bar{\psi}_m(x) \psi_n(y) = \delta(x-y)$,
so Eq.~\eqref{eq:jacobian} involves the divergent quantity $\delta(0)$
and it is clear that this trace will require regularization.

In general, Eq.~\eqref{eq:jacobian} represents a complicated
redefinition of each local operator in the Lagrangian.
However, in a mean field approximation where
$\phi = \phi_0 + \delta \phi$,
its value can be calculated perturbatively in $\delta \phi$.
This approximation is reasonable in the physical situations of interest
for optical phenomena associated with coupling of the scalar to gauge
fields, including those in astrophysics and laboratory experiments.
Working to first order in $\delta \phi$ and discarding an infinite
$\delta\phi$-independent prefactor, we find
\begin{equation}
	\left| \frac{\partial( a, \bar{b} )}
		{\partial( c, \bar{d} )}
	\right|
	\propto
	\exp \tr \frac{3 \alpha}{2} \int \d^4 x \, \sqrt{-g} \;
	\delta \phi \; ( \bar{\psi}_m \psi_n + \bar{\psi}_n \psi_m ) ,
	\label{eq:pre-fujikawa}
\end{equation}
where we have defined a coefficient $\alpha$, satisfying
\begin{equation}
	\alpha = \left. \frac{\d \ln B}{\d \phi} \right|_{\phi_0} ,
\end{equation}
which has dimension $[\text{mass}]^{-1}$.
In what follows it will sometimes be useful to identify
this mass scale explicitly, writing
$M_\alpha = \alpha^{-1}$.

Eq.~\eqref{eq:pre-fujikawa} can be evaluated using
a standard regularization method introduced by Fujikawa
\cite{Fujikawa:1979ay, *Fujikawa:1980eg}.
We write the trace over basis indices in~\eqref{eq:pre-fujikawa}
as $\delta(x-x)$, and make the substitution
\begin{equation}
	\delta(x-x) \rightarrow
	\Tr \exp \Big( \frac{\slashed{\D}_x^2}{\mu^2} \Big)
	\int \frac{\d^4 k}{(2\pi)^4} \; \e{\im k \cdot (x-y)}
	\Big|_{y \rightarrow x} .
	\label{eq:fujikawa}
\end{equation}
The trace is over spinor indices, and
the exponential is to be interpreted
in a matrix sense.
In this expression we have specialized to flat spacetime,
so that $\slashed{\D} = \gamma^a \D_a$ where $\D_a$ is the
flat gauge-covariant derivative. This is the only case which has yet
been required in phenomenology, although the metric coupling could
be reintroduced using $\slashed{e}^\mu \D_\mu$ if desired.
The subscript `$x$' on $\slashed{\D}$ indicates that the differential
operator applies only to the explicit $x$-dependence, and it is
understood that the limit $y \rightarrow x$ is to be taken
\emph{after} allowing $\slashed{\D}_x$ to act on terms
to its right. For finite $\mu$,
Eq.~\eqref{eq:fujikawa} softly
suppresses the contribution of ultraviolet modes with
$k \gg \mu$, while maintaining
gauge-invariance: this is the key virtue of Fujikawa's method.
At the end of the calculation
one removes the regulator by taking
$\mu \rightarrow \infty$.

The operator $\slashed{\D}^2$ can be written
\begin{equation}
	\slashed{\D}^2 = \D^2 - \frac{\im e}{4} [ \gamma^a, \gamma^b ] F_{ab} .
\end{equation}
We write
the Jacobian as a perturbative shift of the action, $\delta S$,
\begin{equation}
	\left| \frac{\partial( a, \bar{b} )}
		{\partial( c, \bar{d} )}
	\right|
	= \e{\im \delta S} .
\end{equation}
Making use of the elementary identity
$\e{-\im k \cdot x} f(\partial_a) \e{\im k \cdot x} = f( \partial_a
+ \im k_a)$, rescaling the momentum integration so
that $k_a \rightarrow k_a/\mu$, and expanding the exponential, we find
\begin{equation}
\begin{split}
	\im \delta S = &\ {-3}\alpha \mu^4 \int \d^4 x \; 
	\delta \phi \int \frac{\d^4 k}{(2\pi)^4} \; \e{-k^2}
	\\ & \mbox{} \times
	\left( 4 - \frac{e^2}{32\mu^4} \Tr
		[ \gamma^a, \gamma^b ] [ \gamma^c, \gamma^d ] F_{ab} F_{cd}
		+ \cdots
	\right) ,
\end{split}
\label{eq:action-shift}
\end{equation}
where `$\cdots$' denotes
operators of lower engineering dimension which
we have not written explicitly,
but which vanish on taking the trace, together with terms suppressed by
more powers of $1/\mu$ which will not contribute in the limit
$\mu \rightarrow \infty$.

The first term in round brackets
$( \cdots )$ in Eq.~\eqref{eq:action-shift} is independent of
$F_{ab}$. It is an infinite renormalization of the
Einstein-frame cosmological constant. This is just a restatement of
the usual cosmological constant problem.
Therefore, discarding this term is harmless.
In a more general theory, we would find infinite renormalizations of
each relevant local operator.
The remaining term is finite.
It is a new contribution which we must absorb in the Einstein frame action.
To evaluate it we Wick rotate to Euclidean signature,
making the substitution $k_0 \rightarrow - \im k_E$,
where $q_0$ is the timelike component of the four-vector $k_a$,
$k_E$ is its Euclidean counterpart, and the sign is fixed by
the requirement that the Euclidean action becomes a Boltzmann weight
of the correct sign.
Carrying out the trace over Dirac indices and
assembling all numerical coefficients, we find
\begin{equation}
	\delta S \supseteq
	\frac{3 e^2 \alpha}{16 \pi^2} \int \d^4 x \; 
	\delta \phi \, F^{ab} F_{ab} .
\end{equation}
In the language of Kaplunovsky \& Louis, this is the renormalization of the
Wilsonian gauge coupling \cite{Kaplunovsky:1994fg}.

A contribution of this form arises for each fermion species.
In a theory with $\Nf$ species of fermion,
the total shift in the Einstein frame action
can be written as a new, local dimension-five operator
coupling $\delta \phi$ and the gauge-kinetic term,
\begin{equation}
	\delta \LagE \supseteq
	\frac{\delta \phi}{M_5} F^{ab} F_{ab} ,
	\label{eq:dim-five}
\end{equation}
where the mass-scale $M_5$ satisfies
\begin{equation}
	M_5 = \frac{16\pi^2}{3 e^2 \Nf} M_\alpha .
	\label{eq:mass-scale}
\end{equation}
Although we have given the calculation only for a massless fermion
and Abelian gauge field, it applies for a non-Abelian field after
straightforward modifications.
Mass terms for the fermions were discussed by
Arkani-Hamed \& Murayama \cite{ArkaniHamed:1997mj},
and can be incorporated perturbatively in the present framework.
Contributions of order $\delta \phi^2$ and higher could be retained,
if desired, by retaining higher-order
terms in the Taylor expansion of $B$
in Eq.~\eqref{eq:pre-fujikawa}
and subsequent expressions.

\section{Discussion}
\label{sec:discuss}

Eqs.~\eqref{eq:dim-five}--\eqref{eq:mass-scale} represent a purely quantum
effect associated with canonical normalization of
the charged fermion species.
Before changing variables,
there is no interaction involving $\phi$ and the gauge field alone;
the only interaction arises from the fermion kinetic term
$\Lag \supseteq B^2 \bar{\lambda} (\slashed{e}^\mu \D_\mu) \lambda
+ \text{h.c}$. After expanding around the mean field $\phi_0$, this
gives rise to
$\phi A \bar{\lambda} \lambda$
vertices involving an arbitrary number of $\phi$
quanta, a single gauge field, and a fermion/antifermion pair.

Changing variables to canonically normalized fields introduces
a new dimension-five operator, Eq.~\eqref{eq:dim-five}, which
yields a contact interaction between $\phi$
and two quanta of the gauge field.
This eliminates the $\phi A \bar{\lambda} \lambda$ vertex.
If the fermion has a mass term, however, then conformal
invariance is broken
and a $\phi \bar{\lambda} \lambda$ interaction persists even
after canonical normalization.
By computing triangle diagrams, one can show that
a contact interaction of the form~\eqref{eq:dim-five} is introduced
at energies sufficiently small that the fermion $\lambda$ cannot be
resolved.
This contribution enters with
the same sign as~\eqref{eq:dim-five},
and its associated mass scale is
\cite{Goldberger:2007zk, *[{the same result was later obtained
independently in }][{}]Brax:2009ey}
\begin{equation}
	M_5' = \frac{48 \pi^2}{e^2 \Nf^{>}} M_\alpha ,
	\label{eq:threshold}
\end{equation}
where $\Nf^{>}$ indicates the number of charged fermion species whose
mass thresholds have been passed.
These heavy fermions do not appear in an effective theory valid at
energies below the threshold;
Eq.~\eqref{eq:threshold} summarizes their residual influence.
We note that, in principle, the mass scale $M_\alpha$ which appears
in~\eqref{eq:mass-scale} and~\eqref{eq:threshold}
can be species-dependent
but in a minimal theory we can expect approximately the same scale
for each species.

Kaplunovsky \& Louis gave an interpretation of
the $\phi$-dependent gauge coupling
as a one-loop effect,
in which
Eq.~\eqref{eq:dim-five} arises from diagrams
with $\lambda$-quanta circulating in the interior of the loops.
However, we emphasize that Eq.~\eqref{eq:dim-five}
must be absorbed into the Einstein-frame action whether or not the
massive fermion species are integrated out.
Therefore, although these two effects are related they are not the same.
The first is a consequence of the non-invariance of the fermionic path
integral measure under a Weyl rescaling. The second is
due to the path integral over fermions whose masses explicitly
break Weyl invariance.

What are the consequences for an effective theory describing the interactions
of $\phi$ with gauge fields at low energies? Such a theory
governs the relevance of $\phi$ for optical interactions in astrophysics
or the laboratory.
Eqs.~\eqref{eq:dim-five}, \eqref{eq:mass-scale}
and~\eqref{eq:threshold} show that when studying the
Einstein-frame phenomenology
of a theory with $\Nf$ total fermion species,
at energies below the mass threshold of $\Nf^>$ of these,
we should augment whatever bare contact term of the form~\eqref{eq:dim-five}
exists by a shift of
$1/M_5''$, where the mass scale $M_5''$ is determined by
\begin{equation}
	M_5''
	=
	\frac{16\pi^2}{e^2}
	\frac{M_\alpha}{3\Nf + \frac{1}{3} \Nf^>} 
	\rightarrow
	\frac{24 \pi^2}{5 e^2 \Nf} M_\alpha ,
\end{equation}
and the limit on the right-hand side corresponds to very low energies,
where all fermion species are integrated out.
If the bare contribution corresponds to a very large mass scale, this
shift will dominate the resulting coupling.
Typically one would expect a coupling to be generated at least by
loops of virtual gravitons, although presumably strongly suppressed.
For a large number of species, we find that the coupling scale to matter $M_m$ can be much larger than $M''_5$.

\section{Conclusion}
We have shown that, in a theory where a scalar field $\phi$ couples to
fermionic matter charged under a gauge symmetry,
a coupling between $\phi$ and the gauge fields is
automatically generated via the noninvariance of the path integral
measure under rescaling of the fermion fields.
This opens up a rich phenomenology
previously associated with axion physics.
Since scalar fields couple to the gauge field strength and can be subject
to the chameleon mechanism, physical predictions will differ between models,
as discussed
in Refs.~\cite{Brax:2007ak,
*Brax:2007hi, *Burrage:2007ew, *Burrage:2008ii, *Burrage:2009mj,
*Schelpe:2010he, Brax:2010jk, Davis:2009vk, *Davis:2010nj}.

\begin{acknowledgments}
	CB is supported by the German Science Foundation (DFG) under
	the Collaborative Research Centre (SFB) 676 and by the SNF.
	ACD was supported in part by STFC
	and the Perimeter Institute of Theoretical Physics.
	DS was supported by the Science and Technology Facilities Council
	[grant number ST/F002858/1].
	AW acknowledges support from NRF, South Africa.
	AW and DS were supported by the Centre for Theoretical
	Cosmology, Cambridge, during the early part of this work.
	We would like to thank Andreas Ringwald for drawing our attention
	to Ref.~\cite{Kaplunovsky:1994fg}
	and Mark Goodsell for suggesting the relevance of
	Ref.~\cite{ArkaniHamed:1997mj}.
\end{acknowledgments}

\bibliography{anomaly}

\end{document}